\documentstyle{article}

\def\QQc{\renewcommand{\baselinestretch}{1.5}}
\def\bq{\begin{quotation}}
\def\eq{\end{quotation}}
\def\beq{\begin{equation}}
\def\eeq{\end{equation}}

\begin{document}
\QQc

\newcommand {\sdot}{\!\!\cdot\!}
\begin{titlepage}
\begin{flushright}
\large
BNL-48160\\
\vspace{0.05in}
ITP-SB-92-54
\end{flushright}
\huge
\vspace{0.22in}
\addtocounter{footnote}{1}
\renewcommand{\thefootnote}{\fnsymbol{footnote}}
\begin{center}
Probing CP Violation Via Higgs Decays to Four Leptons
\vspace{0.4in}
\QQc
\large
\baselineskip=18pt

A. Soni\footnote{\normalsize soni@bnlcl1.bnl.gov}\\
Department of Physics\\
Brookhaven National Laboratory\\
Upton, NY 11793\\

\vspace{.15in}

R. M. Xu\footnote{\normalsize rmxu@max.physics.sunysb.edu}\\
Institute for Theoretical Physics\\
State University of New York at Stony Brook\\
Stony Brook, NY 11794-3840\\

\baselineskip=24pt

\vspace{0.35in}
Abstract
\end{center}
\large
Since decays to four leptons is
widely considered a promising way to search
for the Higgs particle, we show how the same final state can also
be used to search for signals of CP nonconservation.
Energy asymmetries and triple correlations are related to parameters
in the underlying CP violating effective interaction at the
$H^0$-$W$-$W$ and $H^0$-$Z$-$Z$ vertex. Expected size of the effects are
shown to be small for both the Sandard Model and its extension with an
extra Higgs doublet.\\

\vfill
\end{titlepage}
\large

The search for the Higgs particle is clearly one of the top
priorities in particle physics. In view of the remarkable
successes of the Standard Model (SM), some manifestation of the
Higgs boson should exist, responsible for the spontaneously breaking
of the gauge group $SU(2)_L\times U(1)$ down to $U(1)$.
Another important problem in particle physics is the origin
of CP violation. In the SM, CP violation is accounted for by the
Kobayashi-Maskawa (KM) phase \cite{km}. However, there is
considerable interest in searching for sources of CP-violation
other than the KM phase. For example, it is generally believed that
the KM mechanism alone cannot produce sufficient baryon asymmetry
in the Universe \cite{bsw}. A number of extensions of the SM have CP
violation other than the KM phase. Tests of CP nonconservation will
therefore test theories beyond the SM.

Once the Higgs particle is discovered, its properties will have to be
investigated vigorously. In particular its role in CP nonconservation
will undoubtedly get a close scrutiny. Since a very promising
way to search for the Higgs is through its decay to four leptons,
in this paper we show how that final state can also, simultaneously,
be used to study the CP properties of the Higgs particle.
We will thus consider the following processes:
(I) $H^0\rightarrow Z^* Z^*\rightarrow \ell^+\ell^- \bar{\nu}\nu$;
(II) $H^0\rightarrow W^* W^*\rightarrow \ell^+\nu\ell'^- \bar{\nu}'$;
(III) $H^0\rightarrow Z^* Z^*\rightarrow \ell^+\ell^-\ell'^+\ell'^-$
\cite{com1}. Here $\ell$ and $\ell'$ stand for either
the electron or the muon,
and $\nu$ and $\nu'$ stand for any species of neutrino allowed by lepton
flavor conservation and $W^*$ and $Z^*$ can be either on- or off-shell.
Even though we will keep our formalism general, we are primarily
interested in the heavy Higgs mass region ($m_H \ge 2m_W$) in this
paper since the corresponding branching fractions are larger.
Also these processes are expected to have considerable  background
problems in the intermediate mass region ($m_Z \le m_H \le 2m_W$)
\cite{b}.
In the heavy mass region the branching fractions are
approximately $8\times 10^{-3}$, $2\times 10^{-2}$ and $10^{-3}$
\cite{ehlq}, respectively, for processes (I)-(III) where we have summed
over electron and muon final states.

CP-odd observables that we will discuss are the energy asymmetries
and the CP-odd angular correlations
of the charged leptons.
Indeed these are the only CP-odd
observables that can be constructed for the above processes if we assume
that polarizations of the final state leptons are not observed.
There are several different energy asymmetries that can be defined.
The first one is:
\begin{equation}
\delta_1={\frac {<E_+-E_->}{<E_++E_->}}
\end{equation}
This is suitable for
process (I), where $E_{\pm}$ stands for the energy of positively
and negatively charged leptons
this paper). Strictly speaking, the energy asymmetry for process (II)
is meaningful only when $\ell^+$ is the antiparticle of $\ell'^-$. However,
assuming lepton universality and setting all lepton masses to zero,
we can also define a kind of ``flavor-blind'' energy asymmetry.
Thus $\delta_2$, the energy asymmetry for process (II), is
unambiguously defined, in analogy to eq.(1).
Two independent energy asymmetries can also be defined
for process (III), i.e.
\begin{equation}
\delta_3={\frac {<E_+-E_->}{<E_++E_->}}~~~\mbox{and}~~~
\delta'_3={\frac {<E_++E'_+-E_--E'_->}{<E_++E'_++E_-+E'_->}}.
\end{equation}
We recall that
for any of the energy
asymmetries to receive a nonvanishing contribution the amplitude for
the process must have an absorptive
part as required by CPT invariance
since the energy asymmetry is a CP-odd but naive T-even object \cite{vs}.

CP noninvariance can also show up in the angular correlation of
the two decay planes defined by the momenta of the final state leptons.
This is a straightforward generalization of Yang's parity test \cite{cny}.
In the case of only four particles in the final state, all their
momenta need to be tracked down to determine the angular correlations.
Thus the CP-odd angular correlation is only useful in the process (III), i.e.
$H^0\rightarrow \ell'^+ \ell'^- \ell^+ \ell^- $ .
The angular correlation between the
decay planes can be parameterized as \cite{dn}
\begin{equation}
{\frac {d \Gamma}{d \phi}}= {\frac {\Gamma}{2\pi}}\left[
1+\lambda_1 \cos \phi +\lambda_2 \cos 2\phi +\lambda_3 \sin \phi
+\lambda_4 \sin 2\phi \right],
\end{equation}
where $\Gamma$ is the partial decay width for process (III) and
$\phi$ is the angle between the two decay planes.
A nonvanishing $\lambda_1$ indicates parity is violated, whereas
nonvanishing $\lambda_3$ and/or $\lambda_4$ are indications
of CP nonconservation. CP violating
angular correlation effects will be completely
washed out for two identical lepton pairs in the final state, since
there is ambiguity in identifying $\phi$ and $-\phi$. However,
both parity and CP violating effects can be picked up if two lepton
pairs are different, thus we will only consider non-identical lepton
pairs in process (III) when we consider angular correlations.

It is easy to show that $\lambda_3$ is related to the usual
CP-odd triple product correlation. Let $\vec{p}_1$ and $\vec{p}_2$
be momenta of one pair of leptons coupled to one $Z$-boson
and $\vec{k}_1$ and $\vec{k}_2$ be momenta
of the other pair of leptons. A CP-odd triple product $\Omega$ can be
defined as:
\begin{equation}
\Omega\equiv (\vec{p}_1 - \vec{p}_2)\cdot (\vec{k}_1 \times \vec{k}_2).
\end{equation}
In the rest frame of Higgs, $\Omega$ is the only independent CP odd triple
product that can be formed by using lepton momenta only. Then a
measure of CP asymmetry $A_{CP}$ defined as
\begin{equation}
A_{CP}\equiv{\frac {\Gamma(\Omega>0)-\Gamma(\Omega<0)}
{\Gamma(\Omega>0)+\Gamma(\Omega<0)}}
\end{equation}
is just $2\lambda_3/\pi$.

To simplify the calculations we will consistently neglect
lepton masses. Then the most general tensor structure
of the $H^0VV$
($V$ stands for either the $Z$ or the $W$) vertex
relevant for decays to massless leptons
assumes the form
\begin{equation}
m_V\left[ \varrho_V g_{\mu \nu}+\sqrt{2}G_F\varsigma_V
(q_1\cdot q_2 g_{\mu\nu}-q_{1\nu}q_{2\mu})+\sqrt{2}G_F\vartheta_V
\epsilon_{\mu\nu\alpha\beta}q_1^{\alpha}q_2^{\beta}\right].
\end{equation}
Here $m_V$ stands for the mass of
$Z$ or $W$; $G_F$ is the Fermi constant, $q_1$ is the momentum of one of
the vector bosons coupled to the lepton current $j_1^{\mu}$; and $q_2$ is the
momentum of the other vector boson coupled to the lepton current $j_2^{\nu}$.
We have inserted $m_V$ and $\sqrt{2}G_F$ in eq.(6) to make
$\varrho_V,~\varsigma_V$ and $\vartheta_V$ dimensionless.
Coefficients $\varrho_V,~\varsigma_V,~\vartheta_V$ are
functions of $q_1^2$ and $q_2^2$ in general. The origins of different
terms in eq.(6) can be identified as follows. The first term is
the familiar $H^0VV$ tree level coupling accompanying the Higgs
mechanism for giving masses to the gauge bosons.
The second term is from the dimension-5 operator $H^0 FF$
 ($F$ being the field strength of the vector filed $V$)
which can be generated by ``integrating out''
heavier particles in the theory. The last term is from another dimension-5
operator $H^0 F\tilde{F}$, $\tilde{F}$ being the conjugate of $F$.
Note that CP is violated if both $\varsigma_V$ and $\vartheta_V$ are
simultaneously present since $FF$ and $F\tilde{F}$ have opposite
transformation properties under CP. $\vartheta_V$ and $\varrho_V$ cannot
coexist either if CP were a good symmetry\cite{rr}.

In all the calculations in this paper, only the first term in eq.(6),
i.e. the tree level coupling, and its interference with the third term
in eq.(6) will be kept. The reason
for this is that we are interested in numerically retaining only the most
significant terms that reflect CP violation. It is not difficult to see
that this procedure is justified for our purpose because
both $\varsigma_V$ and $\vartheta_V$ in eq.(6) are radiatively induced, and
$\varsigma_V$ is unimportant because it conserves CP and we are only
interested in CP violating effects in this paper.

First we discuss the energy asymmetries. It is straightforward to
show that
$$\delta_1=\delta_3=\delta'_3=-{\frac {8c_Vc_A}{c_V^2+c_A^2}}\delta_Z,~~~
\mbox{and}~~~\delta_2=4\delta_W,$$
where $c_V=-1+4\sin^2\theta_W$ and $c_A=1$ are respectively the vector
and the axial-vector coupling constants of the $Z$-boson to
the charged-leptons;
$\delta_Z$ and $\delta_W$ are defined as \cite{com2},
\begin{equation}
\delta_V=\sqrt{2}G_F{\frac {\int dq_1^2 dq_2^2 \varrho_V \mbox{Im}\vartheta_V
|\vec{q}|^3 q_1^2 q_2^2 \Delta_V}{\int dq_1^2 dq_2^2 \varrho_V^2
|\vec{q}|(3q_1^2 q_2^2+m_H^2|\vec{q}|^2)\Delta_V}}.
\end{equation}
In eq.(7),
$|\vec{q}|$ is the magnitude of the spatial momentum of either of
the gauge bosons in the rest frame of the Higgs boson,
\begin{equation}
|\vec{q}|={\frac {\sqrt{m_H^4+q_1^4+q_2^4-2(m_H^2 q_1^2+m_H^2 q_2^2
+q_1^2 q_2^2)}}{2m_H}},
\end{equation}
where $m_H$ is the Higgs mass; $\Delta_V$ is from the propagators of the
two gauge bosons,
$$\Delta_V = {\frac {1}{[(q_1^2-m_V^2)^2+m_V^2\Gamma_V^2]
[(q_2^2-m_V^2)^2+m_V^2\Gamma_V^2]}},$$
$\Gamma_V$ being the total width of the gauge boson $V$.
The integrations in both the numerator and the denominator of eq.(7)
are over the region $\sqrt{q_1^2}+\sqrt{q_2^2}<m_H$.

In general, $\vartheta_V$ is a function
of both $q_1^2$ and $q_2^2$. However, we will assume $\vartheta_V$ to be a
constant here. This approximation is justified if $\vartheta_V$ is a
slowly varying function of $q_1^2$ and $q_2^2$. Furthermore, the integrands
in eq.(7) are peaked in the region where either of the propagators
can be on-shell. As we have indicated earlier, we are mostly interested in
the case
when both the vector bosons are on-shell since the branching ratios are
larger. Thus it is not a bad approximation to replace the function
$\vartheta_V$ by its value at $q_1^2$ and/or $q_2^2$ set equal to $m_V^2$.
Using the narrow width approximation for both vector bosons, eq.(7) becomes
\begin{equation}
\delta_V={\frac {\mbox{Im}\vartheta_V}{\varrho_V}}
{\frac {\sqrt{2}G_F(m_H^2-4m_V^2)}
{12+{\frac {m_H^2}{m_V^2}}[{\frac {m_H^2}{m_V^2}}-4]}}.
\end{equation}
Both the exact result eq.(7) and the on-shell approximation
eq.(9) for $\delta_1$ and $\delta_2$ are plotted in Fig. 1. From the
figure we see that $\delta_1$ is about
$10^{-3}\times{\frac {\mbox{Im}\vartheta_Z}{\varrho_Z}}$
and $\delta_2$ is about
$10^{-2}\times{\frac {\mbox{Im}\vartheta_W}{\varrho_W}}$.

Now we discuss the CP-odd angular correlations. It is straightforward
to show that the differential partial decay rate is
\begin{eqnarray}
{\frac {d^3\Gamma}{dq_1^2 dq_2^2 d \phi}}&=&{\frac {(c_V^2+c_A^2)^2
m_Z^2|\vec{q}|\Delta_Z}{128(2\pi)^6 m_H^2 }}\left\{ \varrho_Z^2
\left[ \right. \right.
{\frac {8}{3}}q_1^2 q_2^2+{\frac {8}{9}}m_H^2|\vec{q}|^2 + \nonumber \\
&&{\frac {\pi^2(c_V c_A)^2}{(c_V^2+c_A^2)^2}} q_1\cdot q_2
\sqrt{q_1^2 q_2^2} \cos \phi +{\frac {4}{9}}q_1^2 q_2^2 \cos 2\phi
\left. \right]-\sqrt{2}G_F\varrho_Z\mbox{Re}(\vartheta_Z)m_H
|\vec{q}|\nonumber \\
&&\left[ \right.
{\frac {\pi^2(c_V c_A)^2}{(c_V^2+c_A^2)^2}}q_1\cdot q_2
\sqrt{q_1^2 q_2^2}\sin \phi+
{\frac {8}{9}}q_1^2 q_2^2\sin 2\phi \left. \left. \right]\right\}.
\end{eqnarray}
It is obvious from the above equation that the parity violating
coefficient $\lambda_1$ is suppressed by a factor of
$\pi^2(c_V c_A)^2/(c_V^2+c_A^2)^2\sim 10^{-2}$ (see also Ref.\cite{dn,pwy}).
In addition to this suppression, $\lambda_3$ is further
suppressed by the ratio $\mbox{Re}(\vartheta_Z)/\varrho_Z$ rendering
$\lambda_3$ extremely small. The single differential decay
rate with respect to $\phi$ can be obtained by
numerically integrating eq.(10).
$\lambda_3$ and $\lambda_4$ are plotted in Fig. 2, again,
we have assumed that $\varrho_V$ and $\vartheta_V$ are constants.
The units in Fig. 2  are taken to be
${\frac {\mbox{Re}\vartheta_Z}{\varrho_Z}}$. From the
figure we see that $\lambda_3$ stays approximately constant
for $m_H>200GeV$ at $6\times 10^{-4}$; $\lambda_4$ peaks around
$m_H=190GeV$, it varies between $8\times10^{-4}$ and $8\times 10^{-3}$.

In the SM, $H^0$ is a scalar particle. It couples to $Z$- and $W$-bosons
at tree level, with $\varrho_Z=g/\cos\theta_W$ and $\varrho_W=g$.
$HFF$ is induced at one loop level, whereas CP violating interaction
$HF\tilde{F}$ does not arise till two-loop order for $W$-bosons and at
three-loops for $Z$-bosons.
In addition to the suppression by powers of $4\pi$ associated with these
loops, CP violation in the SM will necessarily involve product of small
mixing angles and also perhaps small ratio of masses. It is thus clear that
in the SM $\vartheta_V$ and consequently the asymmetry parameters
discussed above are all expected to be extremely small.

On the other hand, many extensions of the SM have other sources of CP
violation besides the KM phase which may enhance the rate of CP violating
decay of the Higgs particle significantly. As an illustration
we will consider a two-Higgs-doublet
model with softly symmetry breaking term \cite{br}.
The relevant Yukawa coupling in such a model is taken to be
\begin{equation}
-{\cal L}_Y=\lambda_{ij}\bar{Q}^i_L\tilde{\phi}_2 U^j_R+
g_{ij}\bar{Q}^i_L\phi_1 D^j_R+h.c.
\end{equation}
There are three spin-0 neutral bosons $\varphi_1^0$, $\varphi_2^0$
and $\varphi_3^0$ in the theory. If there are no CP violation from the
scalar sector, two of them will be CP-even and the other one is CP-odd.
In the presence of CP violation, they mix through their mass matrix.
Let us call the three neutral mass eigenstates $H_1^0$, $H_2^0$ and $H_3^0$,
then
\begin{equation}
H_i^0=O_{ij}\varphi_j.
\end{equation}
Where $O_{ij}$ are the matrix elements of an orthogonal matrix that
diagonalizes he  mass matrix. $\varrho_Z$ and $\varrho_W$ arise from the
tree-level coupling of $H_i^0$ to two $Z$ and $W$-bosons, while
coefficients $\vartheta_Z$ and $\vartheta_W$ arise through the top quark
loop shown in Fig. 3. It is easy to show that
\begin{equation}
\left({\frac {\vartheta_Z}{\varrho_Z}}\right)_i
={\frac {3m_t^2\kappa_i}{(4\pi)^2 }}
\int^1_0dx \int^{1-x}_0 dy {\frac {2\tilde{c}_a^2(x+y)+\tilde{c}_v^2
-\tilde{c}_a^2}{q_1^2(x^2-x)+q_2^2(y^2-y)-2xyq_1\sdot q_2+m_t^2}},
\end{equation}
\begin{equation}
\left({\frac {\vartheta_W}{\varrho_W}}\right)_i
={\frac {12m_t^2\kappa_i}{(4\pi)^2 }}
\int^1_0dx \int^{1-x}_0 dy {\frac {(x+y)}
{q_1^2(x^2-x)+q_2^2(y^2-y)-2xyq_1\sdot q_2+m_t^2}}
\end{equation}
where $\tilde{c}_a=-1$ and $\tilde{c}_v=1-8/3\sin^2\theta_W$ are axial-vector
and vector coupling constants of $Z$ to top-quark;
$\kappa_i=\cot\beta O_{i3}/O_{i1}$, where $\cot\beta=|v_1|/|v_2|$,
$|v_2|$ and $|v_1|$ are vacuum expectation values of $\phi_2$ and $\phi_1$
respectively. In Fig. 4 we show the both the real and the imaginary part of
${\frac {\vartheta_Z}{\varrho_Z}}$ and ${\frac {\vartheta_W}{\varrho_W}}$
for on-shell $W's$ and $Z's$, with $m_t=130GeV$.

It is clear that both eq.(13) and eq.(14) will develope an imaginary part
only when $m_H>2m_t$. Thus, within the context of the two-Higgs-doublet
model with soft symmetry breaking, energy asymmetry is a useful observable
only when $m_H>2m_t$. Nevertheless, the angular correlation is
useful both for $m_H>2m_t$ and also for $m_H<2m_t$.
In this model, $\kappa_i$ is completely undetermined, although one may
invoke some naturalness argument that it should not be too different from
order one. Assuming $\kappa_i$ is of order one, then we see that typically
$\delta_1\sim 10^{-5}$ and $\delta_2\sim 10^{-4}$; $\lambda_3\sim 10^{-6}$
and $\lambda_4 \sim 10^{-5}$.
Thus in this model process (II) appears the most promising. Indeed, from
Figs. 1 and 4 we see that $\delta_2$ can be as large as
$1.5\times 10^{-3}$  for $m_H$ about 300GeV.
Assuming that the branching ratio ($BR$) for this process is about
$3\times 10^{-2}$, as it is in the SM, we see that the number
of Higgs needed to see such an asymmetry,
given roughly by $(\delta_2^2 \times BR)^{-1}$,
is about $10^7$. This number is about a factor of forty larger than
the expected number of Higgs at the hadron
supercolliders, based again on the SM \cite{ehlq}.
Consequently, at least in this extension of the Standard
Model the resulting asymmetries appear too small to be observable.
On the other hand, there is a large uncertainty in these estimates
as many of the relevant parameters in these models have not been pinned
down. Furthermore, our analysis is completely general so it may be useful
to study what other extensions of the SM will yield for these
asymmetries.  The virtue of these tests of CP
nonconservation is that they can be done at little or no extra cost.

\noindent{\bf Acknowledgement}: R.M.X. would like to thank Jack Smith and
Chung Kao for helpful discussions. The work of R.M.X. is supported in part
by NSF Grant No. PHY 9108054 and that of A.S. is supported in part by
USDOE contract number DE-AC02-76CH0016.

\newpage
\begin{center}
\Large
{\bf Figure Captions}\\
\end{center}
\begin{description}
\item[Figure 1] $\delta_1$ from eq.(7) (solid line)
and from the on-shell approximation (dotted line) in the units of
${\frac {\mbox{Im}\vartheta_Z}{\varrho_Z}}$. $\delta_2$ from eq.(7)
(dashed line) and the on-shell result (dash-dotted line)
in the units of ${\frac {\mbox{Im}\vartheta_W}{\varrho_W}}$.
\item[Figure 2] $\lambda_3$ (solid line) and $\lambda_4$ (dashed line)
in the units of ${\frac {\mbox{Re}\vartheta_Z}{\varrho_Z}}$.
\item[Figure 3] One-loop Feynman diagram contributing to $\vartheta_V$ in
the two-Higgs-doublet model.
\item[Figure 4] Real (solid line) and imaginary (dotted line) part of
$({\frac {\vartheta_Z}{\varrho_Z}})_i$, and real (dashed line) and imaginary
(dash-dotted line) part of $({\frac {\vartheta_W}{\varrho_W}})_i$.
They are all in the units of $\kappa_i$ and $m_t$ is taken to be $130$ GeV.
\end{description}

\begin{thebibliography}{20}
\bibitem{com1}Processes with $Z^*\gamma^*$ and $\gamma^*\gamma^*$
intermediate states may also be considered.
These could be important for relatively light Higgs which are not
the primary focus in this paper.
\bibitem{com2}Eq. (7) is not true for process (III) when the two lepton pairs
are identical, because the interference term is not taken into account here.
However, when we consider the on-shell case, i.e. eq. (9), interference effect
is expected to be minimal.
This is based on the corresponding discussion in $\pi^0$ and $K_L$ decays to
four leptons. In this regard see T. Miyazaki and E. Takasugi,
Phys. Rev. {\bf D8} (1973) 2051.
\end{thebibliography}
\end{document}